\documentclass[preprint,epsfig,aps,showpacs]{revtex4}

\usepackage{epsfig}

\newcommand{\BV}{\left(\begin{array}{c}}
\newcommand{\EV}{\end{array}\right)}
\newcommand{\BM}{\left(\begin{array}{cc}}

\newcommand{\beq}{\begin{equation}}               
\newcommand{\eeq}{\end{equation}}                 
\newcommand{\bqry}{\begin{eqnarray}}              
\newcommand{\eqry}{\end{eqnarray}}                
\newcommand{\bqryn}{\begin{eqnarray*}}            
\newcommand{\eqryn}{\end{eqnarray*}}              

\begin{document}

\setcounter{page}{0}

\title{Valence Quark Distribution in A=3 Nuclei}
\author{C.\, J. Benesh}
\affiliation{Department of Chemistry and Physics, 
Wesleyan College, Macon, GA 31204}

\author{T. Goldman}
\affiliation{Theoretical Division, MS-B283, 
Los Alamos National Laboratory, Los Alamos, NM 87545}
\author{G.\, J. Stephenson, Jr.}
\affiliation{Department of Physics and Astronomy, 
University of New Mexico, Albuquerque, NM 87131}

\date{\today}

\begin{flushright}
\vspace{-1.5in}
{LA-UR-03-4587}\\
\vspace{-0.15in}
{nucl-th/0307038}\\
\vspace*{0.2in}
\end{flushright}

\begin{abstract} 
We calculate the quark distribution function for $^{3}$He/$^{3}$H 
in a relativistic quark model of nuclear structure which adequately
reproduces the nucleon approximation, nuclear binding energies, and
nuclear sizes for small nuclei. The results show a clear distortion
from the quark distribution function for individual nucleons (EMC
effect) arising dominantly from a combination of recoil and quark
tunneling effects. Antisymmetrization (Pauli) effects are found to be
small due to limited spatial overlaps. We compare our predictions with
a published parameterization of the nuclear valence quark distributions 
and find significant agreement.
\end{abstract}

\pacs{13.60.Hb, 21.45.+v, 27.10.+h, 25.30.-c}
\maketitle

\section{Introduction}
The validity of the nucleon approximation -- the approximation 
that atomic nuclei are accurately viewed as bound agglomerations 
of protons and neutrons -- was recognized as essential to nuclear 
physics from its very beginning\cite{weissk}. The development 
of the meson-exchange picture of hadronic interactions did not 
impact this question due to difficulties with field theory 
self-consistencies, (e.g., physical interpretation of off-shell 
quantities,) although the distortion in the nuclear medium of 
the meson (or at least pion) cloud surrounding a nucleon was also 
recognized\cite{RhoBrown}. The validity of the approximation remained 
assumed after the advent of Quantum ChromoDynamics (QCD) allowed 
dynamical examination of the quark and gluon substructure of nucleons, 
despite the natural question as to why these strongly interacting, 
composite objects do not distort each other's internal structure 
beyond recognition when brought into close proximity\cite{McLerran,Piet}, 
as in a nucleus.

It was therefore disturbing when the EMC collaboration first 
showed that the structure function of a nucleus containing A 
nucleons, as measured in deep inelastic lepton scattering (DIS) 
is not a simple multiple (A) of the corresponding structure 
function of a free nucleon\cite{EMC}, although it was clear 
that recoil effects (as in a Fermi gas model, see Ref.\cite{Bodek}) 
would produce non-zero response beyond the kinematic limit of 
Bjorken-x ($x_{Bj} =1$) for a nucleon with concommitant 
implications for the $x_{Bj} \le 1$ region.  The experimental
result has become steadily clearer with time\cite{review}. A 
number of heuristic approaches (nucleon swelling, binding energy 
rescaling, meson convolution models) produce reasonable agreement 
with some of the data\cite{Close} without providing any deep, 
dynamical understanding. Indeed, Miller\cite{Gerry} has recently 
remarked that the situation remains unresolved in terms of a 
satisfactory understanding of the phenomenon: Nothing short of 
changing the structure of nucleons in the nuclear medium seems 
to be called for. Such a change occurs completely naturally in 
the quark nuclear model used here\cite{A=4}, and may be compared 
with the work of others\cite{Hood,OTHRS} who have considered 
the contributions to the EMC effect from quark Pauli effects (the 
minimal, quantum mechanically required change in the structure of 
nucleons in the nuclear medium).

We have approached the problem from a different direction, building 
on a description of nuclei directly in terms of relativistic, 
four-component Dirac quark wavefunctions\cite{A=4}. The physics of 
that model
provides an immediate, qualitatively clear, justification 
of the nucleon
approximation: The tendency for quark wavefunctions to
reduce their kinetic energy by spreading out between nucleons in close
proximity (delocalization) is countered by a reduction, engendered by
this delocalization, in the amount of the attractive interaction
energy due to the color magnetic hyperfine interaction  between the 
correlated quarks in a nucleon\cite{dibaryon}. Detailed quantitative
calculation confirms this insight\cite{A=4,A=3}, and shows that the
probability of finding a quark in a location other than what would be
expected if its wavefunction were unaltered from that of an isolated
nucleon is limited to the level of a few percent.

The results of the variational calculations in the model also 
reproduce qualitatively (and systematically) correct binding 
energies in the A=3 and A=4 systems\cite{footnote}, as well as 
overall matter distributions consistent with those experimentally 
inferred from low energy electron scattering and other 
measures\cite{rms}. To be explicit, we obtain a binding energy 
of $(20~\pm~4)$~MeV and $rms$ matter (not charge) size of
1.34~fm for $^{4}$He\cite{A=4} (cf., 1.42 fm in an ab initio
calculation using a realistic nucleon-nucleon potential\cite{JAC}) 
and a binding energy of $(4~\pm~2)$~MeV for $^{3}$He/$^{3}$H\cite{A=3} 
with a slightly smaller $rms$ size. Although not comparable to shell 
model accuracies, and still not as good as the results of the ab initio
calculations using pairwise nucleon interaction potentials\cite{Carlson} 
noted above, these results are obtained with no free parameters beyond
those in the quark model used to fit the nucleon and $\Delta$-baryon
masses, and represent a fraction of a percent accuracy in the energy
calculated per quark.

All of these features depend on the model introduction of a
geometrically complex, mean-field describing the overall confining
potential encountered by a quark in a nucleus. The potential is 
composed of individual, linearly confining (nucleon-like) potential 
wells, set out in a regular array, (equilateral triangle for $A = 3$
and tetrahedron for $A = 4$,) in a body-fixed frame, separated by a 
variationally determined distance scale, $d$, (common side length) 
and truncated on the midplanes between each pair of wells. The 
structure of this `egg-crate' potential, if extended to larger A, 
may be relevant to the observations of Cook and others\cite{Cook}.

The quark delocalization is described by the variationally determined
parameter, $\epsilon$, which is, roughly speaking, the probability
amplitude for a quark to appear in the interior of a nucleon-like well 
other than the one in which it originates. We find a value of 
$\epsilon~=~0.136$ for $^{4}$He and $\epsilon~=~0.104$ for 
$^{3}$He/$^{3}$H. The corresponding values of the distance scale are 
$d~=~1.75$~fm for $^{4}$He and $d~=~1.80$~fm for $^{3}$He/$^{3}$H.

Both of these parameters are jointly determined variationally by
minimizing the overall (one body kinetic and potential plus two-body 
color magnetic interaction) energy of the quark nuclear configuration 
in each nucleus. Thus, this approach allows for no new free parameters 
and predicts the quark nuclear structure within the assumptions of the
model and approximations made to carry out the calculations.

In the next section, we describe the structure of the wavefunctions 
for the model as applied to $^{3}$He/$^{3}$H. In Sec.\ref{calc} we 
describe the method used for calculating the quark distribution 
functions from those wavefunctions, including the effects of Pauli 
antisymmetrization. We present our results in Sec.\ref{results} and 
discuss those results in Sec.\ref{disc}, comparing them with a 
phenomenological extraction of the valence quark distributions in 
nuclei from Ref.\cite{Kumano}, and concluding with a review of 
additional elements not yet included in the model.

\section{Wavefunctions}
\label{wvfcns}

The variational wavefunctions of the model are built on full, four 
component, Dirac solutions of the quark wavefunctions in an isolated, 
Lorentz scalar, confining potential well\cite{A=4}. The Lorentz 
character of confinement remains a subject of dispute\cite{Olsson}, 
but there is little difference between the spatial wavefunctions found 
for linear confining potentials of either scalar or vector character. 
If the absence of spin-orbit effects in the hadron spectrum\cite{Isgur} 
does reflect the character of the confining potential\cite{Ginocchio}, 
then an equal admixture is to be expected, further weakening the 
distinction with the scalar presumption. The potential used is 
\beq
V_{c}(r) = k(r-r_{0}) \label{potl}
\eeq
where $k = 0.9$ GeV/fm is the conventional slope and $r_{0} = 0.57$ fm 
has been chosen to remain within the error band of phenomenological fits 
to spectral data\cite{Tye}. The negative region near the origin is expected 
to account, albeit crudely, for the (undoubtedly vector) color Coulomb 
potential at short distances. More modern analyses that explicitly 
separate out the short distance color Coulomb component do not markedly 
affect the value of the slope used here. 

The basic wavefunction, $\psi_{0}(\vec{r})$, is found by solving the
Dirac equation for the potential in Eq.(\ref{potl}) for a massless
(negligible mass) quark. The quarks are thus `current' (not constituent) 
$u$ and $d$ quarks so that there are no complications in determining 
the photon coupling to them needed for our deep inelastic scattering 
(DIS) calculations below.

The variational spatial wavefunction, $\psi_{j}(\vec{r},\epsilon)$, 
for each quark in the nucleus, is composed as
\beq
\psi_{j}(\vec{r},\epsilon) = \frac{1}{N(\epsilon)} [ \psi_{0}(\vec{r} - 
\vec{R_{j}}) + \epsilon \sum_{i \neq j} \psi_{0}(\vec{r} - \vec{R_{i}})]
\label{qkwfcn}
\eeq
where the color and flavor indices have been suppressed and 
$N(\epsilon)$ is a normalization factor. The quantities 
$\vec{R_{i}}$ describe the three origins of potentials, each 
of the form $V_{c}(r)$, but truncated on the midplanes between 
each $(i,j)$ pair, corresponding to the average location of a 
nucleon in a body-fixed frame for $^{3}$He/$^{3}$H. By symmetry, 
these average locations define an equilateral triangle for the 
minimum energy configuration of the nucleus as a whole. The 
scale of this triangle is given by the variational parameter, 
$d$, which has a value of $d = 1.80$ fm, while the other 
variational parameter attains the value  $\epsilon = 0.104$ 
at the variational minimum. 

The three quarks with spatial wavefunctions $\psi_{j}(\vec{r},\epsilon)$ 
are antisymmetrized in color and appropriately coupled in spin-flavor to
nucleon quantum numbers. These are, in turn, antisymmetrized over the
three well locations. Beyond this, the final wavefunction includes all
antisymmetrization (pairwise, triples, etc.) between all nine quarks.
(Note that these ``nucleons" are {\em not} sharply localized; that is 
true only for the origins of the potentials, so that the geometrically 
complex, confining, mean field potential is well defined.)

In order to perform the calculations of quark distributions 
that follow, the wavefunction is written in Foch-space as
\beq
\vert A> = {\cal N}\int d^3r_{CM} \int d\Omega\, \chi^{abc} 
N_a^\dagger(\vec R_1) N_b^\dagger(\vec R_2) N_c^\dagger(\vec R_3)
\vert EB \{R_i\}>,
\eeq
with ${\cal N}$ a normalization factor and $\chi^{abc}$ the nuclear 
spin-isospin wavefunction. The effective creation operator for a 
nucleon of spin-isospin $a$ is given by 
\beq
N_a^\dagger(\vec R_i) = T_a^{\alpha\beta\gamma}b^\dagger_{0\alpha}(\vec R_i)
b^\dagger_{0\beta}(\vec R_i)b^\dagger_{0\gamma}(\vec R_i)
\eeq
where $b^\dagger_{0\alpha}(\vec R_i)$ creates a quark of 
spin-isospin-color $\alpha$ with wavefunction $\psi_0(\vec r-\vec R_i)$ 
centered on the i-th well, and $T_a^{\alpha\beta\gamma}$ is the quark 
spin-isospin-color wavefunction for a nucleon of spin-isospin $a$.

In $\vert A>$, the integration over $\vec r_{CM}=\frac{1}{3}\sum
\vec R_i$, the center of the triangle, projects onto a state of 
zero momentum, while the integration over the three Euler angles
denoted by $d\Omega$ projects onto zero orbital angular momentum. 
Finally, for non-zero values of $\epsilon$, the individual quark 
creation operators are replaced by 
\beq
b^\dagger_\alpha(\vec R_i,\epsilon) = b^\dagger_{0\alpha}(\vec R_i) 
+\epsilon\sum_{j\neq i}  b^\dagger_{0\alpha}(\vec R_j)
\eeq

It is useful to note that the wavefunction contains no D wave 
components, and that the wavefunction is completely antisymmetrized 
at both the quark and nucleon levels. As a result, quarks within 
a given ``nucleon'' are indistinguishable from one another, and, 
as a result of the summation over the well locations and  
integrations over nuclear orientations, the individual ``nucleons'' 
are also indistinguishable from one another.

\section{Calculation}
\label{calc}

The calculation of valence quark distributions here parallels  
previous work on quark distributions in the nucleon\cite{9} and 
for the color-hypefine-induced flavor distortions between $u$ and 
$d$ valence distributions\cite{CJB}. The valence distributions 
are calculated at a low momentum scale $\le$ 1 GeV$^2$, but one 
still sufficiently large that it is reasonable to conceive of a 
nucleon/nucleus
as a simple object that may be described by a 
quark model. At this scale, the twist-two contribution to the 
structure functions of the target are projected out by taking the 
Bjorken limit on the momentum transfer.

 For unpolarized scattering, the relevant matrix elements for a 
quark distributions in nucleus A are given by\cite{8}
\bqry
q_{\alpha}(x) &=& \, {1\over 4\pi} \int  d\xi^-  \, e^{iq^{+}\xi^-}
   <A\vert\bar\psi_{\alpha}(\xi^-)\gamma^+\psi_{\alpha}(0)
   \vert A>\vert_{LC} \nonumber\\
\bar q_{\alpha}(x) &=&-{1\over 4\pi} \int  d\xi^- \, e^{iq^+\xi^-} 
   <A\vert\bar\psi_{\alpha}(0)\gamma^+\psi_{\alpha}(\xi^-)\vert A>\vert_{LC}, 
\label{xq}
\eqry
where $q^+=-Mx/\sqrt{2}$ (with $x\equiv x_{Bj}$ the Bjorken scaling
variable), $\psi_{\alpha}(\bar\psi_{\alpha})$ are field operators for 
quarks of flavor ${\alpha}$, $\gamma^+$ is the light cone projected 
($0~+~3$ component) Dirac gamma matrix, and the subscript LC indicates 
a light cone condition on $\xi$, namely that $\xi^+=\vec\xi_\perp=0$.

The approach we adopt consists of a straightforward evaluation of the 
matrix elements of Eq.(\ref{xq}) in a Peierls-Yoccoz projected momentum 
eigenstate, assuming that the time dependence of the field operator is 
dominated by the lowest eigenvalue of the Dirac equation used to obtain 
the wavefunctions of the struck quark. The details of this procedure are 
described in Ref.\cite{9}, where the valence quark distribution for flavor
$i$ in a free nucleon are shown to be given by
\beq
xq_V^{i}(x)= {MxN_{i}\over \pi V}\left \{ \left [ 
\int_{\vert k_-\vert}^\infty k dk \, G(k)
\left (t_{0}^2(k) + t_{1}^2(k) + 
2{k_-\over k}t_{0}(k)t_{1}(k)\right ) \right ] 
+\left [ k_- \rightarrow k_+ \right ] \right \},
\eeq
where 
\bqry
G(k) &=& \int \frac{d^3r}{4\pi} e^{i\vec k\cdot\vec r}  
\big{(}\Delta(r)\big{)}^2 EB(r), \nonumber \\
V &=& \int \frac{d^3r}{4\pi}  
\big{(}\Delta(r)\big{)}^3 EB(r),\nonumber\\
t_{0}(k)& =& \int r^2 dr j_0(kr) u(r),\nonumber\\
t_{1}(k)& =& \int r^2 dr j_1(kr) v(r),\nonumber\\
\Delta(r)&=& \int d^3z \psi^\dagger_{0}
({\vec{z}}-{\vec{r}}) \psi_{0}(\vec{z}),\nonumber\\
 \eqry
with $N_i$ the number of valence quarks of flavor $i$ 
in the nucleon, $\psi_{0}({\vec{r}})$ the ground state 
valence quark wavefunction of Eq.(\ref{qkwfcn}), with 
upper and lower components $u(r)$ and \mbox{$i\vec{\sigma}
\cdot\vec{r}v(r)/r$} (both times a fixed spinor), 
respectively, and \mbox{$k_\pm = \omega\pm Mx$}, with 
$\omega$ the ground state struck quark energy eigenvalue. 
Here, and in the nucleus, $\omega$ is taken to be the one 
body quark energy. 

Also, \mbox{$EB(r) \equiv <EB,{\vec{R}_{CM}} = {\vec{r}}~|~EB, 
{\vec{R}_{CM}} = {\vec{0}}>$} is the overlap function for 
two ``empty bags'' (potential wells without the quarks in 
them) separated by a distance $r$, which accounts for the 
dynamics of the confining degrees of freedom. In this paper, 
we assume that the function $EB(r)$ is a constant for both 
the nucleon and the nucleus\cite{12a}. We will discuss below 
some effects of this assumption. 

Note that the function $\Delta(r)$ measures the overlap of two 
(bra and ket) quark wavefunctions centered in wells separated 
by a distance $r$, and that the recoil function $G(k)$ 
measures the probability of finding all of the spectator degrees 
of freedom of the unprojected state carrying a net momentum 
opposite to that of the struck quark. It is also useful to note 
that $G(k)$ is the Fourier transform, over the separation 
between the centers of the struck quark wavefunctions, of the 
spatial overlap of the spectator degrees of freedom in the 
unprojected state.

Remarkably, generalizing this expression to nuclei only 
requires modification of $G(k)$ and the normalization 
volume, $V$. To see this, recall, as we have already noted, 
that in the momentum and angular momentum projected 
wavefunctions of the nucleus, the individual nucleons are 
indistinguishable, so that we may always consider the 
struck quark in the bra and the ket to originate in the 
{\it same} well. Consequently, after a change of integration 
variables from the center of mass position to the separation 
of the centers of the struck wells, all of the complexity of 
the nuclear wavefunction, including exchange effects and 
tunneling, can be incorporated into a redefinition of the 
recoil function and of the requisite normalization volume. 
Explicitly,
\bqry
G(k,\epsilon) = & & \int d\Omega^\prime\int d\Omega \int 
\frac{d^3r}{4\pi} e^{i\vec k\cdot\vec r} \sum_{m=0}^{18}
\Big{(}
\sum_{\{n_{mjk}\}} PF(\{n_{mjk}\}) \epsilon^m \nonumber \\
& & \times \prod_{j,k=1..3} \Big{[} \Delta(\vec R_j-\vec R^\prime_k)
\Big{]}^{n_{mjk}} 
EB(\{\vec R_\alpha,\vec R^\prime_\alpha\},\Omega,\Omega^\prime), \nonumber \\
V(\epsilon) = & & \int d\Omega^\prime\int d\Omega \int 
\frac{d^3r}{4\pi} \Delta(r) \sum_{m=0}^{18}
\Big{(} 
\sum_{\{n_{mjk}\}} PF(\{n_{mjk}\})  \epsilon^m \nonumber \\
& & \times \prod_{j,k=1..3} \Big{[} \Delta(\vec R_{j}-\vec R^{\prime}_k)
\Big{]}^{n_{mjk}} 
EB(\{\vec R_\alpha,\vec R^\prime_\alpha\},\Omega,\Omega^\prime)). \nonumber \\
\eqry
In this expression, the integer $m$ indicates the total number 
of quarks in the bra and ket that have tunneled. Since there 
are many possible configurations with the same $m$, it is 
necessary to characterize each configuration by a set of nine 
integers $n_{mjk}$ which indicates the number of quarks from 
well $j$ in the bra which are contracted with quarks from well 
$k$ in the ket.  The factor $PF(\{n_{mjk}\})$ is the 
Permutation-Flavor-Color-Spin (PerFlaCS) factor that goes with 
the exchange/tunneling term associated with $\{n_{mjk}\}$.
Finally, the empty-bag overlaps will, in general, depend on the 
locations of the well-centers in the bra($\{\vec R_\alpha\}$) 
and ket($\{\vec R^\prime_\alpha\}$) and on their relative 
orientations.

The calculation is simplified by the existence of two small 
parameters, $\epsilon\approx\Delta(d)\approx 0.1$, which 
allow for a consistent truncation of the sums appearing in 
$G(k,\epsilon)$. Since at least two quarks must be exchanged 
between different wells, and each such exchange requires an 
overlap between quarks in different wells, we expect  Pauli 
effects are of order $\Delta^2(d) \approx 0.01$ or smaller. 
Further suppression of exchange effects comes from the PerFlaCS 
factors, which are smaller for the exchange terms than for the 
direct term. Similarly, if a single quark tunnels to a different 
well, it must overlap with a quark from the original well, so 
tunneling effects are of order $\epsilon \Delta(d) \approx 0.01$ 
or smaller. 

In contrast to the exchange effects, tunneling is not suppressed 
by a PerFlaCS factor, but rather tends to be enhanced due to 
combinatorics. Our expectation, then, is that quark tunneling 
will play a larger role than exchange effects in modifying the 
shape of the valence distribution. In the following, we keep 
only the leading contributions from the latter. 
 
The dominant effect, analogous to Fermi motion, comes from the 
modification of $G(k)$. Since $G(k)$ measures the probability 
that the spectator quarks in the unprojected state have a total 
momentum equal and opposite to that of the struck quark, the naive 
expectation is that systems with more spectators will tend to have
larger contributions from large momentum states, resulting in an 
enhancement of the valence distribution at large $x$. In the 
absence of the rotational projection, one expects that the overlap 
function $G(k)$ falls off roughly as the Fourier transform of 
$\Delta^{n_{s}}(r)$ where ${n_{s}}$ is the number of spectator 
quarks. In the limit of a large number of spectators, $\Delta^{n_s}(r)$ 
is very sharply peaked, so that $G(k)$ becomes essentially constant 
and, as expected for a large system, the momentum projection is 
replaced by an incoherent average over center of mass positions. 
When additional collective effects, such as rotations, are included, 
the correlations between the motions of quarks in different wells 
is weakened,  yielding spectator overlap functions that approach the 
limiting case more slowly,
as the additional degrees of freedom allow 
the (bra and ket) spectator wells to be closer to each other than the 
struck quark wells are in some orientations of the system.
  
The integrations over orientations were calculated using 
Gauss-Laguerre quadrature over the five independent orientation 
(Euler) angles defined by the two planes of the equilateral 
triangles for the bra and ket state nuclei, and over the 
separation between the $\vec{R_{j}}$, i.e., locations, of 
the origins of the confining potential wells for the struck 
quarks. Exchange terms are calculated to all orders in
$\Delta(d)$ and tunneling terms are included to order 
$\epsilon\Delta(d)$.

We note in passing that the $^{3}$He/$^{3}$H calculation 
carried out here has been additionally checked by restricting 
the calculation to the case of a quasi-deuteron (two wells 
with their `nucleons' coupled to $I=0, J=1$, with values for 
${\epsilon}$ and $d$ similar to those used for $^{3}$He/$^{3}$H, 
even though these do not correspond to realistic values for the 
actual deuteron). The results of our earlier calculation\cite{BAPS} 
of this case, which was carried out with a completely independent 
(and differently structured) code for the integrations, are 
accurately reproduced. 

\section{Results}
\label{results}

We show our d-quark valence distributions in Fig.~1 for $d=$ 
1.8 fm and three values of $\epsilon$, with and without the 
Pauli antisymmetrization corrections. As has been noted above, 
the Pauli effects are negligible for these distributions. This 
conclusion differs from that derived in Ref.\cite{Hood}, but 
is in agreement with Refs.\cite{OTHRS}. In both instances, 
this is due to the smaller overlap we find between quarks 
from different quasi-nucleons than is the case for the 
nuclear wavefunctions used in Ref.\cite{Hood}, which allow 
nucleons to come into closer proximity to one another than is 
the case for more conventional nuclear wavefunctions\cite{OTHRS}. 

\begin{figure*}[t] 
\includegraphics[height=2.0in]{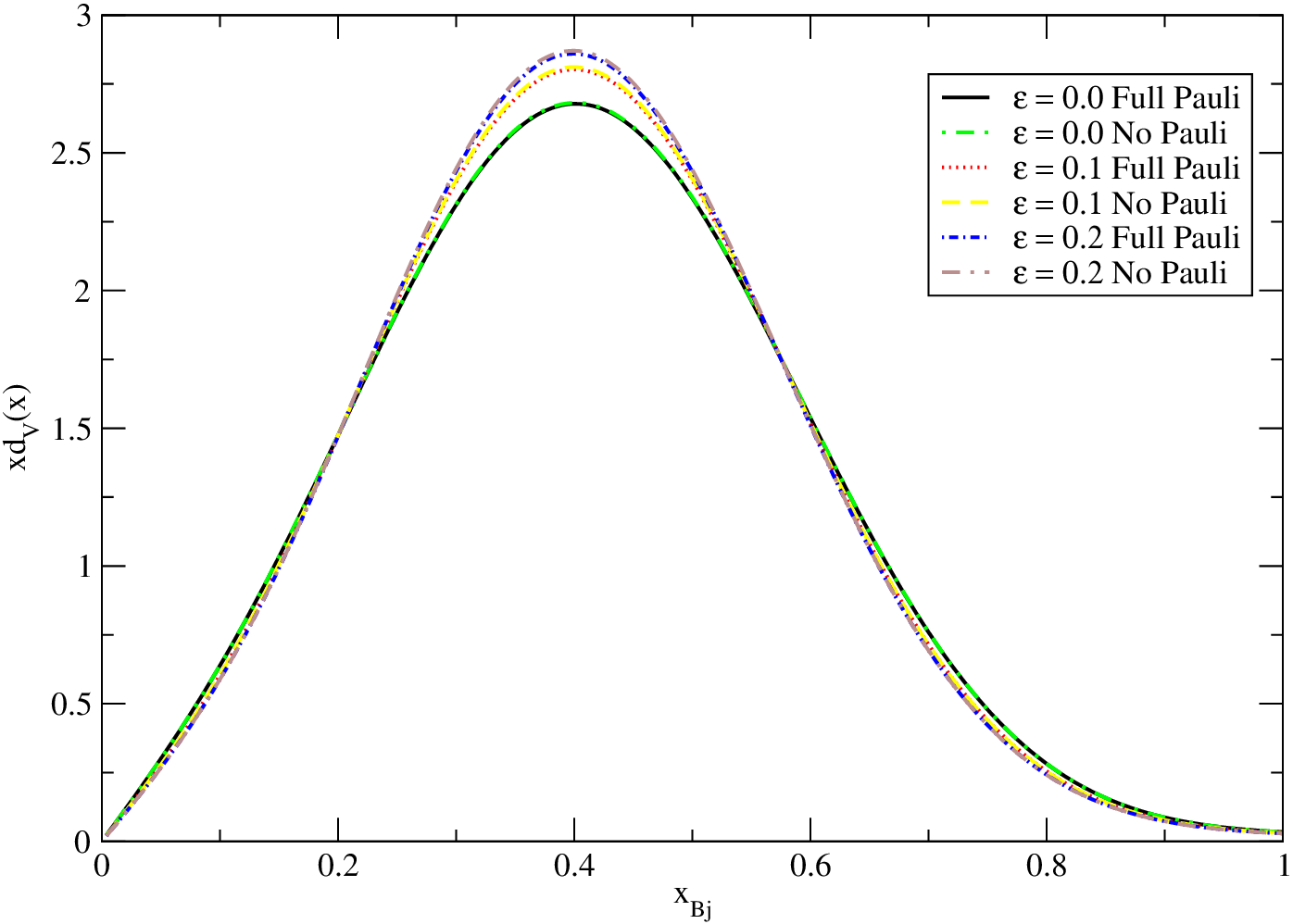} 
\caption{ A=3 d-quark distributions with and without Pauli 
terms for $\epsilon = 0.0, 0.1\, {\rm and}\, 0.2$ and 
$d=$~1.8~fm. See text for an explanation of why the absence 
of discernable differences due to the inclusion of the 
Pauli antisymmetrization is to be expected.} 
\label{xdvaldist} 
\end{figure*} 
	
In our earlier quasi-deuteron calculations\cite{BAPS}, we also 
obtained results that showed a very small effect from Pauli 
statistics -- the results for $I=0, J=1$ and $I=1, J=0$ states 
were very similar to each other. Here we understand the result 
as being due to the fact that, in the absence of tunneling, 
at least two additional powers of the overlap, $\Delta(d)$, 
are required for every pairwise antisymmetrization of a ket 
wavefunction relative to a bra wavefunction and vice versa. 
Since our calculations find that $\Delta(d) \sim 0.1$ at best, 
this leads to, at a minimum, a two order of magnitude suppression 
of quark statistics effects. 

Furthermore, there is an additional suppression due to the 
combination of spin, color and isospin factors associated 
with the Pauli exchanges. Since the spin, color and flavor 
factors do not change for the tunneling terms, and since it 
is possible for a quark to tunnel into the same well as the 
quark with which it is being exchanged,the tunneling and 
exchange terms may partially compensate for each other. Thus, 
since there are simply more possible tunnelings, we expect, 
and find, very small effects from statistics alone, but 
significant effects from tunneling. 

In Fig.~2, the isospin averaged ratio of the valence quark 
distribution in $^3$He to that in a free nucleon is shown 
for $d=$ 1.8 fm and several values of $\epsilon$. For 
comparison, we also show a parametrization of this 
ratio\cite{Kumano} drawn from data. Even in the absence of 
tunneling, we obtain a qualitatively correct EMC effect, due 
to the modification of $G(k)$ caused by the larger number 
of spectators available to share momentum with. Such an effect 
is apparent in even semi-realistic nuclear models, such as a 
Fermi gas picture of a nucleus\cite{Bodek}.  The large $x$ 
behavior has been noted before, and, as has been argued by 
West\cite{GBW}, the normalization constraint on the valence 
distribution then also requires the rise at $x$ near zero. 
Since we have already shown that Pauli effects are negligible, 
the $\epsilon = 0$ distribution may be interpreted as being 
analogous to the effect of Fermi motion within the nucleus.

\begin{figure*}[t] 
\includegraphics[height=2.0in]{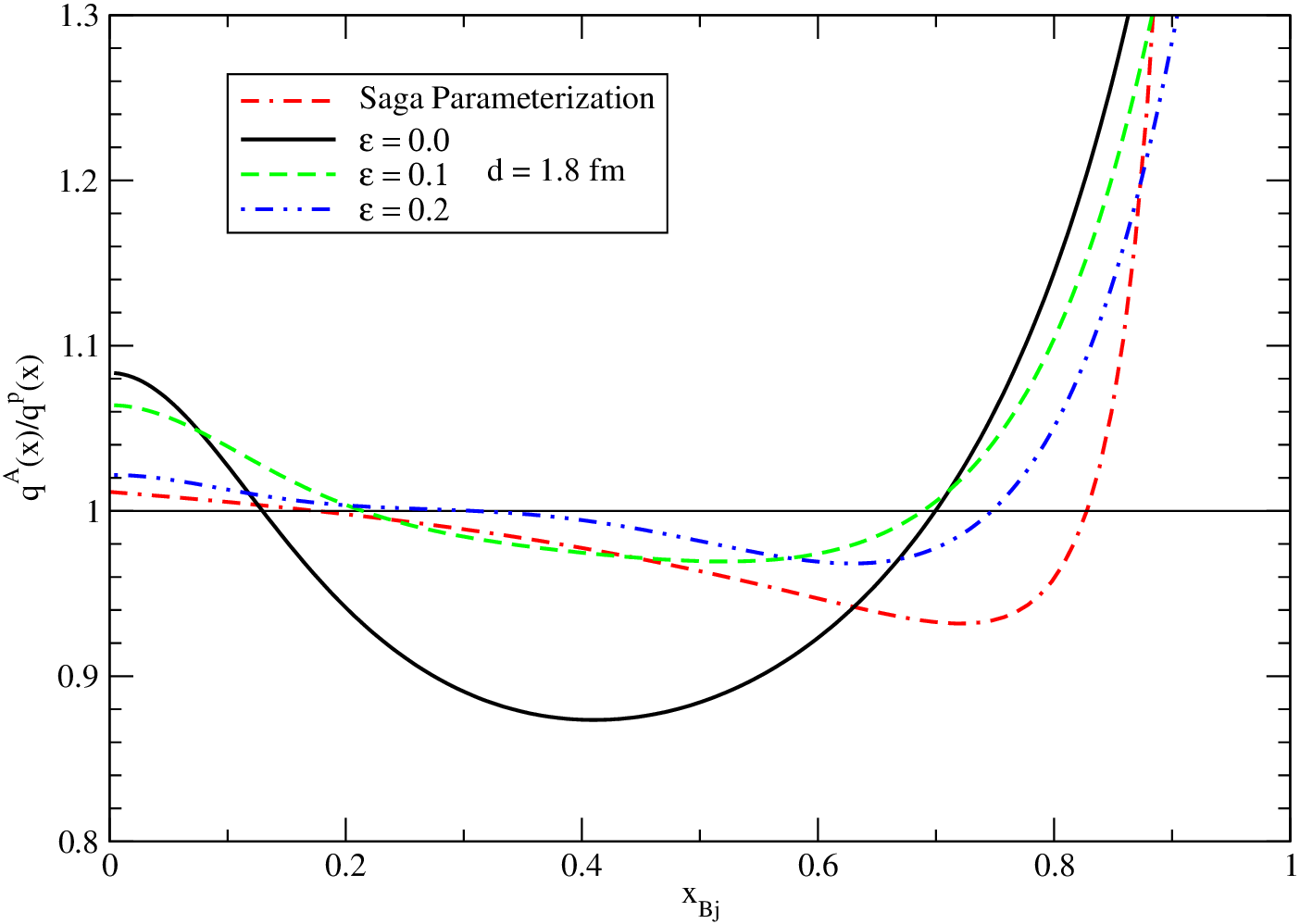} 
\caption{ The $^{3}$He/$^{3}$H EMC ratio for valence quarks for three
values of $\epsilon$ at fixed $d = 1.8$ fm. The parametrization of the
data in Ref.\cite{Kumano} is also plotted for comparison.} 
\label{qdistratio} 
\end{figure*} 

We note also that the $\epsilon=0$ term significantly overpredicts 
the size of the effect, but that the ratio softens considerably 
when the tunneling corrections are included. Physically, this is 
because the quark tunneling terms weaken the correlation between 
the motions of the quarks in the individual wells, enhancing the 
probability of finding quarks with low momenta. We find this result 
to be extremely encouraging, as it indicates that quark tunneling, 
which plays a critical role in producing the correct binding in 
this model, is equally important to the EMC ratio. 

Even with the tunneling terms, however, the calculated ratio still 
rises too rapidly at large $x$. We attribute this to the strong 
correlation between the well centers and consequently the quarks 
within the wells. We have investigated the effect on the nuclear 
wavefunction of including additional, collective oscillations/excitations 
of the well centers; in the case of the deuteron with no 
tunneling, it can be shown analytically that these collective 
motions tend to decrease the valence ratio at large $x$ until 
ultimately the free nucleon response is restored in the limit of 
uncorrelated wells. We expect a similar softening in the $^3$He
ratio when collective effects are included, which will improve the 
agreement with data in both the high and low $x$ regions.

In Fig.~3, the dependence of the ratio of valence distributions 
is shown as a function of the separation of the wells for {\it 
fixed} $\epsilon$. In that case, the ratio is relatively 
insensitive to the distance between the wells. For large separations, 
we see an increase in the enhancement at large $x$. As the separation 
between the wells increases, the overlap of the bra and ket decreases 
more rapidly as their relative orientations change, which in turn 
decreases the coherence of angular projection. Since, as we have 
already noted, the coherent averaging over relative orientations 
softens the valence distribution, the valence distributions harden
as the size of the system increases. For smaller $d$ values, we also 
see an enhancement of the valence distribution at small $x$ and a 
corresponding decrease at larger $x$.   

\begin{figure*}[t] 
\includegraphics[height=2.0in]{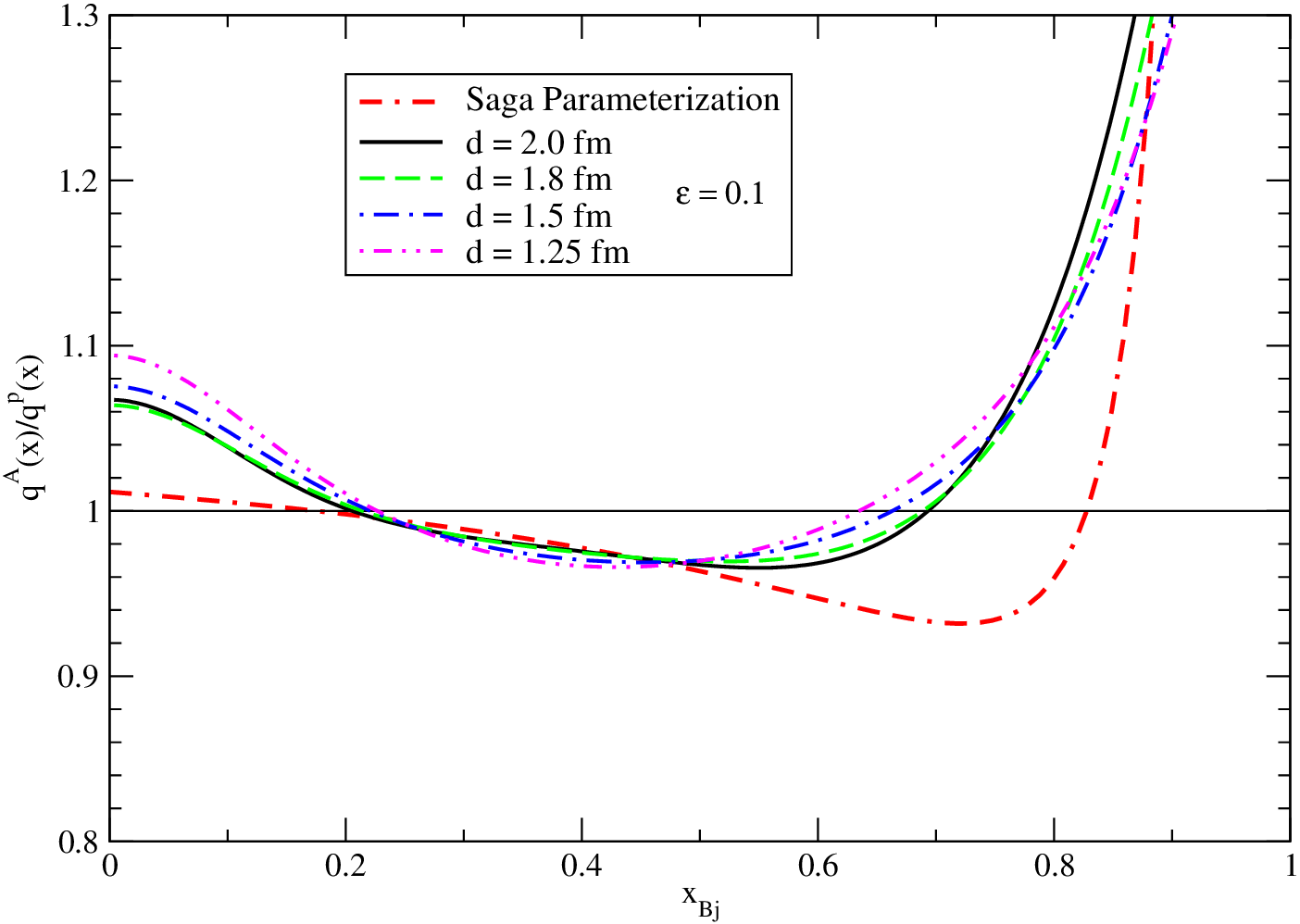} 
\caption{ The $^{3}$He/nucleon valence ratio for for five
values of $d$ at fixed $\epsilon = 0.1$. The parametrization of 
the data in Ref.\cite{Kumano} is also plotted for comparison..} 
\label{qdistratio2} 
\end{figure*} 

Since the values of $d$ and $\epsilon$ are strongly correlated 
in the quark nuclear model\cite{A=3,A=4}, the valence distribution 
ratio should not be viewed as a function of either variable alone. 
In Fig.~4, we show the valence distribution ratio for correlated 
values of $d$ and $\epsilon$ chosen to lie approximately on the 
line of minimum energy in a plot of energy versus $d$ and $\epsilon$ 
obtained from the model. This plot gives a better description 
of how the ratio is expected to change as the separation between 
wells is varied, and, due to the slow motion of the well centers 
compared to quarks, is also relevant for the consideration of 
possible collective effects through a Born-Oppenheimer 
approximation.

\begin{figure*}[t] 
\includegraphics[height=2.0in]{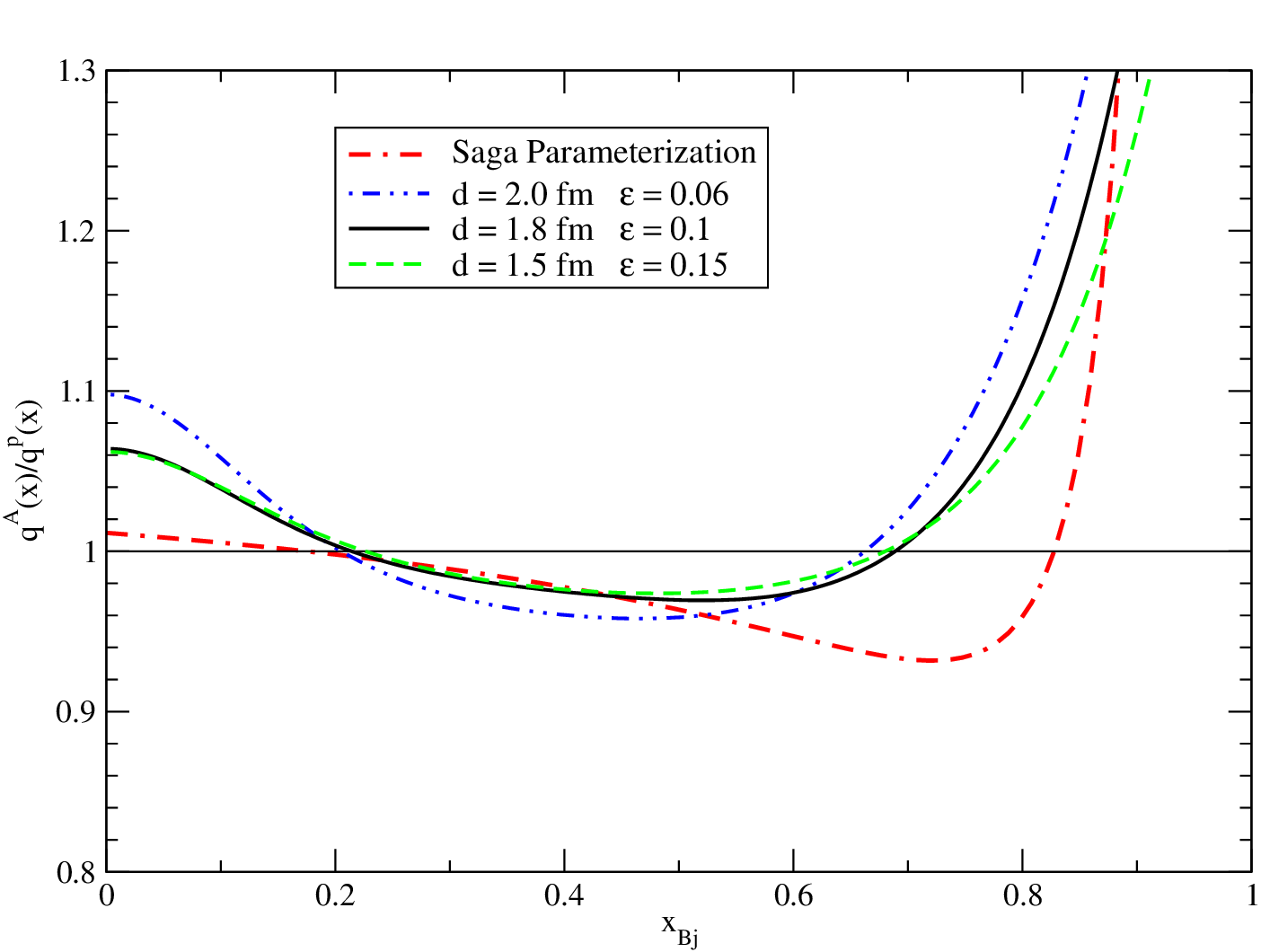} 
\caption{ The $^{3}$He/nucleon valence ratio for correlated
values of $d$ and $\epsilon$ chosen to minimize the energy 
of the nucleus. The parametrization of the data in 
Ref.\cite{Kumano} is again plotted for comparison.} 
\label{qdistratio3} 
\end{figure*} 

\section{Discussion}
\label{disc}

The qualitative agreement of these calculations with the expected 
ratio of valence distributions demonstrates that it is possible 
to formulate a quark-based description of nuclei which correctly 
reproduces both their observed low energy properties (binding 
energy, $rms$ size) and, {\it without introducing any additional 
parameters}, also produces a shift in quark distributions comparable, 
both systematically and in magnitude, to the EMC data. Compellingly, 
the delocalization of quarks produced by tunneling, which plays 
a crucial role in generating phenomenologically suitable binding 
energies, is also critical for producing the right magnitude of 
changes in the valence distributions. 

It should be noted that it is precisely this delocalization 
of quarks which the model employs to provide for the physics 
conventionally described by meson exchanges between nucleons, 
and that it does so without introducing any additional antiquark 
amplitude. The latter, predicted in pion excess models\cite{piex} 
that have attempted to account for the EMC effect based on the 
meson exchange picture of nuclear binding, is inconsistent with 
experimental measurements, using Drell-Yan techniques\cite{E772}, 
of the antiquark amplitude in nuclei. We are, therefore, encouraged 
in the belief that quark tunneling plays an essential role in the 
modification of the structure of the nucleon within a nucleus and 
the concommitant nuclear binding. 

Despite the qualitative succcess of the model as described here, 
there are a number of avenues for quantitative improvements and 
extensions that should be pursued:
\begin{itemize}

\item{} Collective Degrees of Freedom

In the current incarnation of the model, the positions of the 
potential wells are dictated by a rigid geometrical structure 
which may be rotated and translated, but neither the size nor 
the shape of this structure varies. This strong correlation 
between the well positions, and consequently between the quark 
wavefunctions defined with respect to them, leads to strong 
enhancement of the valence distribution at high $x$, overpredicting 
the data in that region. Allowing the centers of the wells to 
oscillate around their equilibrium positions decreases this 
correlation, which leads to softer distributions more consistent 
with the data and to an overall picture of the nucleus that is 
in better accord with standard nuclear physics. (In the case of 
the quasi-deuteron, where the calculation is simpler, we have 
shown analytically that allowing the wells to move relative to 
one another softens the quark distributions, ultimately recovering 
the free nucleon distribution in the free particle limit.)

\item{} Convergence of the $\epsilon-\Delta$ expansion.

In the present calculation, we have included only the leading 
contributions in $\epsilon$. Since $\epsilon\approx 0.1$, and 
since most tunnelings produce additional off-diagonal overlap 
factors ($\Delta(d)$'s), our expectation is that the calculation 
converges rapidly. In the case of the quasi-deuteron 
calculations, where it is possible to calculate all possible 
tunnelings, we have explicitly verified that this is a good 
approximation. Due to increasing combinatoric factors, the 
validity of this approximation is less certain here, and the 
question of higher order corrections should be explored.
  
\item{} Self-consistency of the confining potential

Here, the confining potential has been fixed by fiat, rather 
than determined by a self-consistent generation due to the 
quark density distribution. We have in mind an analog of the 
view that confinement is related to the formation of a gluonic 
condensate in the QCD vacuum, whereas the perturbative vacuum 
is (at least partially) restored in the presence of a 
non-negligible quark matter density. In the tunneling regions 
between quasi-nucleons, the density is somewhat higher than 
that near the (fuzzy) matter surface of an isolated nucleon, 
which will further reduce resistance to tunneling. Still not 
investigated is the question of whether iteration along these 
lines leads to convergence or to complete breakdown of the 
barrier. If the latter were to occur, the justification found 
in the model for the nucleon approximation would disappear.

A dynamically generated potential also allows relaxtion of the 
assumption that the function $EB(r)$ is a constant. This will 
in turn produce softer free nucleon quark distributions that 
would again improve agreement with those observed experimentally. 
Additionally, such a model allows for calculation of changes in 
the parton distribution due to changes in the confining degrees 
of freedom.    

\item{}Contributions to the binding from long range pion 
exchange between quarks 

In a non-relativistic version of this quark nuclear model, 
the introduction of pion exchange between pairs of quarks 
beyond a minimum separation (short distance cutoff) has 
only small effects on compact objects and is only significant 
for the delicate binding of the deuteron\cite{footnote}. 
Nonetheless, we expect inclusion of this physics to have 
some effect on the precise values of $d$ and $\epsilon$.

\item{} Large A Nuclei

As the number of spectator degrees of freeedom in the system 
grows, the number of wavefunction overlap factors, $\Delta(r)$, 
grows, so that the total overlap of the system becomes more 
sharply peaked in coordinate space. Eventually, this trend will 
``lock down'' the center of mass and rotational degrees of 
freedom due to increasingly stringent alignment requirements 
on the bra and ket states. A comparison of ``locked'' bra and 
ket configurations with the calculated result would be of 
considerable interest. Such locking would considerably simplify 
calculations for large A nuclei and invite consideration of the 
possibility of effects analogous to those in ``powder diffraction'' 
X-ray experiments on solids being observable in DIS.	
\end{itemize}

While these items may well be significant amendments to the model 
and the calculational result, they appear to us to be both of an 
offsetting character among themselves, and qualitatively similar 
to the effects already examined. We are therefore strongly encouraged 
by the qualitative similarity to the observed EMC effect, in size and 
shape, already evidenced in our model calculation. 

\section{Acknowledgments}
This work was supported in part by the National Science Foundation
under Grant PHY0071658 and in part by the US Department of Energy 
under contract W-7405-ENG-36.

\end{document}